\documentclass[12pt]{iopart}
\begin{document}

\title[On Dynamical Quantization]{On Dynamical Quantization}

\author{H\'ector Calisto\dag\ and\ C. A. Utreras-D\'{\i}az\ddag}
\address{\dag\ Departamento de F\'{\i}sica, Facultad de Ciencias,
Universidad de Tarapac\'a, Casilla 7-D Arica, Chile}
\address{\ddag\ Instituto de F\'\i sica, Facultad de Ciencias,
Universidad Austral de Chile, Casilla 567, Valdivia, Chile}

\ead{hcalisto@uta.cl}

\ead{cutreras@uach.cl}

\begin{abstract}
In this article we review some results obtained from a generalization of quantum
mechanics obtained from modification of the canonical commutation relation
$\left[q,p\right]={\rm i}\hbar$. We present some new results concerning relativistic
generalizations of previous works, and we calculate the energy spectrum of some
simple quantum systems, using the position and momentum operators of this new formalism.
\end{abstract}
\submitto{\EJP}


\maketitle

\section{Historical introduction}
The validity of any physical theory depends on the experimental
data set from which it was originally abstracted. Although a theory may
be well established, is not completely unexpected that it fails, or gives
unsatisfactory results, when it is applied under experimental sufficiently
different conditions that those which originated it.

Newtonian mechanics, for example, contains three basic postulates:

\begin{enumerate}
\item The existence of an absolute space time

\item The particles move throughout well defined trajectories.

\item Space time is continuous.
\end{enumerate}

The first two postulates were abandoned at the beginning of last
century. As far as the continuity postulate, one may ask whether or
not it is a logical necessity, or if it must be accepted for some
fundamental reason. The development of relativity and quantum
mechanics showed that nature can impose constraints to our
measurements; these constraints are related to the existence of two
fundamental constants: the speed of light $c$ and Plank's constant
$h$. Within this same context, we notice that:

\begin{itemize}
\item Heisenberg quantum mechanics, based upon the canonical commutation
relation:
\begin{equation}
\label{Eq01}
\left[ q,p\right] =\mathrm{i}\,\hbar
\end{equation}%
was formulated more than seventy years ago, originating from
experimental data on atomic physics, that is to say, from phenomena
whose characteristic energies range from a few  eV to about $100$
eV. In the particle physics experiments, the involved energies are
in the range from $10^{9}$ eV to $10^{12}$ eV. On the base of this
observation alone, the question arises on whether the same
commutation relations are still valid, or if some suitable
modification or generalization is required. This question has been
reinforced by experimental observations which suggest that, at high
energies, completely new phenomena are observed, which are very
difficult, or perhaps impossible, to explain within the framework of
the usual quantum mechanics. One of these phenomena is the
confinement of quarks; that is to say, the fact that there are
quarks within the hadrons, which cannot be observed as free
particles.

\item Relativistic quantum mechanics takes into account the two
experimental constraints previously mentioned, but uses differential
equations for fields. Since all the physical laws must be
verifiable, at least in principle, these equations imply that it has
to be possible to measure small space and time intervals without any
finite limit. The validity of this assumption has been verified for
distances of up to the order of $10^{-19}$cm ~\cite{Dineykhan}. To
be realistic then, we should consider the value $\ell $ of the
smallest measurable distance, like an empirical parameter to be
determined by experiments, instead of assuming a priori that $\ell =
0$.
\end{itemize}

Later in this work, we will assume that $\ell $ is a universal
constant in all the inertial reference frames, just as $c$ and $h$,
and we shall prove that it is possible to construct a quantum theory with the
constants $c$, $h$ and $\ell $ without falling into logical inconsistencies when
$ \ell \rightarrow 0$.

The continuity assumption enters in the Euclidean geometry through the
postulate of infinite divisibility of any line segment, together with
the famous postulate on parallel lines. Although both postulates are based
on uncontrollable physical extrapolations, they were seen like evident truths,
in the sense that alternative assumptions seemed unacceptable. One  researcher
maintained that the Euclidean postulate on parallel lines could be demonstrated by
reduction to an absurd proposition. Nevertheless, Lobachevsky, Bolyai and Riemann,
between 1829 and 1854, discovered that it is possible to construct logically consistent
non-Euclidean geometries. It is natural to ask, then, what would it happen if one
abandoned continuity postulate.

According to the literature ~\cite{Mees}, around 1870, Clifford
considered a modification of the Newton laws of the movement, but
without changing the other postulates of classical mechanics. He
simply provided the absolute space time with a discrete structure,
and assumed that the particles may only appear and exist at the
points of the resulting network, they would act like lights that can
be ignited and extinguished one after another. This concept of
discontinuous motion reappeared ~\cite{Mees} after the development
of relativity, but now combined with the idea of a maximum velocity
$c$.

Nevertheless, Einstein had modified the physics in a much deeper
form. He recognized that ideally precise measurements of space and
time intervals are subject to a universal constraints, namely, we
may only obtain results that are related one to the other in such a
way that the speed of light in vacuum has the same value $c$, for
any direction and in all the inertial reference frames. Heisenberg
modified this idea, requiring that the motion of
atomic electrons could be described in terms of all the possible
values of a measurement. Using spectroscopy data, he constructed the
matrix mechanics, where the concept of coordinates was generalized
to be in agreement with the old quantization rules. After the
development of the wave mechanics, Heisenberg formulated his famous
incertitude relationships, demonstrating very clearly the existence
of other universal restriction. A particle may only be localized in
space-time with a precision that depends on the incertitude accepted
between the momentum and energy of the particle, when these
observables are defined in a given inertial reference frame
of reference,   by their wave properties, and Planck's constant.

Perhaps the single event that gave form to the
physics of the twentieth century, was the very surprising
discovery that nature can impose constraints to our measurements, a
fact that also modified the status of the physical laws. Instead of
beginning directly with statements about reality, we make statements
about the knowledge that we can obtain from the reality. This
knowledge is the result of measurements that are subject to
universal constraints, which must be included in the formulation of
the physical laws. Relativistic quantum mechanics combines the
effects of $c$ and $h$, but evidence exists today that nature could
impose a third universal constraint on physical laws.

Pauli, in a review of the basic principles of quantum mechanics~\cite{Pauli}
stated that only the relativistic quantum mechanics is logically complete,
and expressed vigorously his belief that new limitations in the
possibilities of measuring would have to be expressed more directly in a
future theory, and that these would be associated with an essential and deep
modification of the basic concepts of the formalism of present quantum
theory. Pauli also held that the concepts of space and time on very
small scales need fundamental modification.

The origin of this fact lies in that the calculated values of some
physical observables become infinity when the continuous theories
are extrapolated to very small distances, although the measured
values are in fact finite. This difficulty first appeared in the
classic theory of electromagnetism. Quantum electrodynamics
attenuated these divergences, but it did not remove them. With the
purpose of controlling these divergences, around $1930$,
Heisenberg~\cite{Heis1,Heis2}, proposed to replace the continuous
space time by a discrete structure. However, at first sigth,
discrete structures break relativistic invariance, a fundamental
requirement of any theory. Later Snyder~\cite{HS1}, suggested the
idea of using a non commutative structure, and showed that this
necessarily implies the existence of a length scale below which the
notion of physical points ceases to exist. Remarkably, in the Snyder
method the space time remains invariant under Lorentz
transformations, and it is becomes possible that when this method is
used in a field theory it would provide an effective cut-off, that
is to say, a minimum length scale in space-time to which the theory
is sensible, eliminating therefore the infinities.

Unfortunately, the theory of Snyder is not invariant with respect to
translations~\cite{Yang}, and after some initial
developments~\cite{HS2}, this idea fell into oblivion, mainly
because the renormalization program was revealed appropriate to
consistently yield finite numerical values for the observable
magnitudes in quantum electrodynamics, without resorting to non
commutativity. Some time later, in the fifties, von Newman
introduced the term noncommutative geometry when discussing about a
geometry in which an algebra of functions is replaced by a
noncommutative algebra~\cite{Gibbs,Madore}. Nevertheless, the first
example of a noncommutative space that clearly was recognized as so
it is the quantum phase space. In fact the first considerations on
their quantified differential geometry were developed by Dirac in
1926~\cite{Dirac1,Dirac2}. In these works, Dirac discovered the
algebraic structure of the quantum phase space, postulating his
celebrated quantization method for classical theory, consisting in
replacing the Poisson bracket of the classic observable by
$\mathrm{i} \,\hbar $ times the commutator of the associated quantum
operators. In this way, the coordinates of the phase space $p$ and
$q$ become non commutative operators, whose commutator is equal to
$\mathrm{i}\,\hbar $. Since these operators do not commute, they
cannot be simultaneously diagonalized and the notion of space
disappears. In other words, the non commutativity of the operators
$p$ and $q$ imply an incertitude relationship between their observed
eigenvalues, which replaces the notion of individual points phase
space; the closestly related idea remaining in quantum theory being
that of the \textit{the Bohr cell}. In the limit $\hbar \rightarrow
0$ recovers the ordinary phase space.

This particular algebra of operators was the one that inspired the
more radical idea of replacing the coordinates $x_{\mu }$ of the
space time by non commutative operators. As it happens in the
previous case, the relationship $[x_{\mu },x_{\nu }]\neq 0$ implies
an incertitude principle  between different coordinates in space
time  that destroys the idea of points at short distances. One can
argue that as the Bohr cell replaces the points of the classic phase
space, the appropriate intuitive notion to replace a point is one
Planck cell of dimensions given by the Planck area.

More recently, the French mathematician Alain Connes developed one more
formal definition of the notion of non commutativity from a mathematical
point of view~\cite{Connes1,Connes2}. For some time the formalism of Connes was
applied to some physical systems, but with very little success, and was subsequently
abandoned  due to this; however, it generated a renewed interest in the ideas of
Snyder about non commutative space time.

Further motivation for non commutative theories comes from the idea
that, in a quantum theory that includes gravity, the nature of space
time must change at distances of the order of the Planck length. The
momentum and the energy required to make a measurement at these
distances, would by itself modify the geometry of space
time~\cite{Witten1}. A way to formulate mathematically this is to
postulate that, on a scale smaller than the Planck length, the space
time is not a differentiable variety, but it has the structure of
non commutative space time. Then, a quantum theory of gravity which
contains or predicts non commutative coordinates, seems to have good
possibilities of being intrinsecally regulated. The string theories
have already suggested from the eighties the possibility of a non
commutative space time~\cite{Witten2} and appears as the main
candidate for a quantum theory of the gravity.

Also, non commutative field theories play an important role in the
area of the condensed matter, which provides not only specific examples of
mathematical models used to explore the properties of space time in the
physics of high energies and quantum theory of gravity, but that represents
specific applications in an area of increasing interest and impact. A
classic example is the electron theory in an external magnetic field,
projected on the lower Landau level, that can be treated like a non
commutative theory. Clear examples of these applications arise in the study of
the quantum Hall effect~\cite{Girvin,Belisard}.

An recent and convincing examnple of a non commutative theory, in
the area of condensed matter, is the quantum theory of mesoscopic
electrical circuits developed by Li and Chen~\cite{Li1,Li2}, that takes
explicit account of the discretization of electric charge, leading
to a new commutation relationship  between the charge and current operators, similarly
to those studied in the physics of high energies and quantum theory of gravity. Several
advances and applications in the context of the mesoscopic circuits with
discrete charge can be found~\cite{Flores1,Flores2}.

In what concerns the purely mathematical aspect, the traditional framework
of geometry and topology is the set of points with some particular
structure that we call space. Nevertheless, as it was discovered very
early, fundamental objects such as elliptical curves are better, not in terms
of the set of points, but in examining the continuous functions that can be
defined on them. Weierstrass opened up a whole new way in geometry when studying
directly the set of complex functions that satisfy an algebra with
particular addition rule, and to derive the set of points from these.

In non commutative geometry, the general concept to replace sets of points
by an algebra of functions is extended. In many cases, the set of points
is completely determined by the algebra of functions, then, the set of
points can be left, and all the information may be obtained from the funtions alone.
 On the other hand, in many cases, the set of points is very complex and a
direct examination does not provide useful information. In such cases, when
the problem is studied from the algebraic point of view, it is common to find
that it contains by itself all the necessary information. Nevertheless, this algebra
is in general noncommutative. Then the process consists of discovering first
how algebras of functions determine the structure of a set of points, and
then to determine which are the relevant properties of these algebras that
do not depend on the commutativity. After doing this, one can to study
noncommutative geometry, generated by an arbitrary noncommutative algebra.
Von Newman was the first in trying to describe such quantum spaces
rigorously calling to its study geometry without points.

The ideas of noncommutative geometries were retaken in the eighties
by the mathematicians Connes and Woronowicz~\cite{Woronowicz}, who
generalized the notion of a differential structure to the
noncommutative case, that is to say, to arbitrary algebras. This was
completed by a definition of a generalized integration, which
provide a more fuller description of the noncommutative space time,
and allowed the definition of field theories in such spaces.

Summarizing, noncommutative theories have been revealed as useful
tools in theoretical physics; they appear as much in the physics of
high energies, for the description from a fundamental level of the
space time on small scale, as in the area of condensed matter to
describe the quantum Hall effect and in the quantum theory of the
mesoscopic electrical circuits. The enormous activity around these
theories mainly is closely related to the appearance of the
noncommutativenes in the limit of low energies of the string theory
mentioned previously. Since the string theory is the only well-known
theory that could unify all the fundamental interactions, it is
possible that the problems of control of divergences in the quantum
theory of fields, and the quantization of gravity are in last term
intimately related by means of some type of noncommutative algebra.

In this work, we begin with a brief review of  the generalization of
quantum mechanics via the of canonical commutation relation, which was
proposed in the eighties by Professor Igor Saavedra~\cite{Saavedra1,Saavedra2,Saavedra3,
Calisto1,Calisto2}. In several dimensions, this theory provides a noncommutative
algebra between the space coordinates and in its relativistic version it
predicts a space time in which the time is a continuous variable, while at the
same time the space has a discrete structure. With few changes we will use this formalism
to calculate energy spectrua of some simple quantum systems.

\section{Dynamical Quantization}

The purpose of this section is to review the generalization of quantum
mechanics through the canonical commutation relations, proposed by
Professor Igor Saavedra and his collaborators in~\cite{Saavedra1,Saavedra2,Saavedra3,
Calisto1,Calisto2}. In professor Saavedra's own words, this generalization was inspired by the
following reasons:

\begin{enumerate}
\item \underline{Aesthetic}: In general theory of relativity, space
is not given a priori; instead, it is given by the energy, the space is curved
in the proximity of a massive star. On the contrary, in the usual quantum
mechanics, the space, represented here by the position variable, is known
beforehand, that is to say, it is independent of the physical phenomena. This
reveals an evident and unsatisfactory asymmetry between the macrouniverse,
described by general relativity, and the microuniverse, described by the
laws of the usual quantum mechanics.

\item \underline{Curiosity}: Only from an intuitive point of view, it is a very 
surprising fact that the same commutation relations extracted from
atomic physics could continue to be valid for energies that are twelve orders
of magnitude greater.

\item \underline{Phenomenology}: It is possible that some phenomena of the
physics of high energies, such as the confinement of quarks and certain regularities 
exhibited by the so-called heavy photons, are connected with the geometry of space.
\end{enumerate}

Then, with the purpose of investigating these questions, Profesor Saavedra
proposed a generalization of the canonical commutation relationship of the form:
\begin{equation}
\label{Eq02}
\left[ q,p\right] =\mathrm{i}\hbar +\frac{\mathrm{i}\ell }{c}\mathcal{F}(q,p) ,
\end{equation}
where $\ell $ is a constant with dimensions of length and $c$ is the speed
of the light in vacuum. In the course of these investigations we
assumed that the momentum operator $p$ is well-known, so that the
commutation relation~(\ref{Eq02}) determines the position operator $q$ when
the function $\mathcal{F}(q,p)$ is given. In the low energy limit, therefore, $q$ is
the usual position operator: $q=\mathrm{i}\hbar \,d/dp$.

The function $\mathcal{F}=\mathcal{F}(q, p)$, in general, depends on
the dynamics of the problem, and therefore, also the operator $q$
and its eigenvalues; that is to say, physical space is not given
here a priori, but it is determined by the physics of the problem
represented by the choice of function $\mathcal{F}$. This is the
origin of \textit{dynamical quantization}.

In addition, it is assumed that a Hamilton function $H=H(q,p)$ exists and
that the Heisenberg equations of motion:
\begin{equation}
\frac{d\Omega}{dt}=\frac{\mathrm{i}}{\hbar}\left[H,\Omega\right],
\end{equation}
for any dynamic variable $\Omega$, continues to be valid.

\subsection{Non relativistic problems}

The simplest choice for $\mathcal{F}$ that takes into account a possible
dependence of the canonical commutation relation with the energy is:
$ \mathcal{F}=H(q,p)$, which leads to the new commutation relation:
\begin{equation}
\left[ q,p\right] =\mathrm{i}\hbar +\frac{\mathrm{i}\ell }{c}H(q,p)
\end{equation}%
and to a new uncertainty principle:
\begin{equation}
\Delta p\Delta q\geq \left\vert \frac{\hbar }{2}+\frac{\ell E}{2\,c}%
\right\vert ,
\end{equation}%
from which we see that the product $\Delta p\Delta q$ grows linearly with the
energy in this approach. This result was verified experimentally, except by
logarithmic corrections by M. Giffon and E. Predazi~\cite{Giffon}, using data from
the physics of high energies. These authors obtained an approximate value for
the parameter $\ell $ of this theory, that is to say: $\ell =2.3\times
10^{-6}$ fm.

For the case of a free particle in one dimension:
\begin{eqnarray}
\left[ q,p\right] &=&\mathrm{i}\hbar +\frac{\mathrm{i}\ell }{2\,m\,c}p^{2}=%
\mathrm{i}\hbar \left( 1+\delta ^{2}p^{2}\right) \\
\delta ^{2} &=&\frac{\ell }{2\,m\,\hbar \,c}
\end{eqnarray}
The corresponding problem eigenvalue problem for the position operator:
\newline
\begin{equation}
q\,\psi (p)=\lambda \,\psi (p)
\end{equation}
lead to the physical space; S. Montecinos, I. Saavedra and O. Kunstmann~\cite{Saavedra3},
found that the spectrum is discrete:
\begin{equation}
\lambda _{n}=2\,n\,\hbar \,\delta ,\quad 0,\pm 1,\pm 2,\ldots .
\end{equation}
Then, the physical space generated by the hamiltonian $H=p^{2}/{2\, m}$, in
one dimension is a lattice in which the minimum length interval is:
\begin{equation}
\Delta q_{\mathrm{min}}=2\,\hbar \,\delta =\sqrt{\frac{2\,\hbar \,\ell }{m\,c}}
\end{equation}
Once the particle has been located in an arbitrary point of the one
dimensional space, the rest of the space \textit{feels} it, that is to say,
the lattice appears; in this sense, geometry acts like a constant force:
a linear potential. In addition, if $\Delta q_{\mathrm{min}}$ is the space
extension of an extended object, it does not make sense to ask for his
constituents since no test particle can go 'within' it.

\section{Relativistic generalization}

Our starting point constitutes the observation that, for a non relativistic
free particle
\[
\mathcal{F}=\frac{p^{2}}{2\,m}=\frac{1}{2\,m}\,p\cdot p,
\]%
which suggests a relativistic generalization of the form:
\begin{equation}
\label{Eq11}
\mathcal{F}=\frac{p_{\mu }\,p_{\nu }}{m}
\end{equation}%
Then, the equations~(\ref{Eq02}) and (\ref{Eq11}) provide the following relativistic
generalization:
\begin{equation}
\label{Eq12}
\left[ q_{\mu },p_{\nu }\right] =-\mathrm{i}\hbar \left( \,g_{\mu \,\nu
}-\delta ^{2}{p_{\mu }\,p_{\nu }}\right)
\end{equation}%
where $g_{0\,0}=-g_{k\,k}=1$ for $k=1,2,3$, $g_{\mu \,\nu }=0$ for $\mu \neq
\nu $ and from now on:
\begin{equation}
\label{Eq13}
\delta ^{2}=\frac{\ell }{m\,\hbar \,c}
\end{equation}%
As before, we will assume that the operators $p_{\mu }$ are known, so
that the equation~(\ref{Eq12}) determines the position operators $q_{\mu }$ in
the four dimensional Minkowski space. The $p_{\mu }$ act as multiplicative
operators which commute among them:
\begin{equation}
\left[ p_{\mu },p_{\nu }\right] =0
\end{equation}%
In this generalization, these not necessarily represent momentum operators,
although in the limit $\delta \rightarrow 0$, we will require that they
recover their usual meaning in quantum mechanics.

In order to determine position operators $q_{\mu }$ who satisfy $({12})$, we
postulate the general form:
\begin{equation}
\label{Eq15}
{q}_{\mu }=F_{\mu \,\alpha }(p_{0},p_{1},p_{2},p_{3})\frac{\partial }{%
\partial p_{\alpha }}+\mathrm{i}\hbar \,\kappa \,p_{\mu }
\end{equation}%
where $F_{\mu \,\alpha }$ is a function to be determined, and we use
the Einstein convention (repeated indices are implicitly summed over),
$\kappa $ is a real constant. Later we shall see that the choice
of $\kappa $ determines a weight function in the definition of the internal product.

Computing the commutator between $p_{\nu }$ and ~(\ref{Eq15}) we obtain:
\begin{equation}
\label{Eq16}
\left[ q_{\mu },p_{\nu }\right] =F_{\mu \,\alpha
}(p_{0},p_{1},p_{2},p_{3})\delta _{\nu \,\alpha }=F_{\mu \,\nu
}(p_{0},p_{1},p_{2},p_{3})
\end{equation}%
Here $\delta _{\nu \,\alpha }$ is the Kronecker delta: Comparing ~(\ref{Eq16}) with
(\ref{Eq12}) we find:
\begin{equation}
F_{\mu \,\nu }(p_{0},p_{1},p_{2},p_{3})=-\mathrm{i}\hbar \left( \,g_{\mu
\,\nu }-\delta ^{2}{p_{\mu }\,p_{\nu }}\right)
\end{equation}%
Then:
\begin{equation}
\label{Eq18}
{q}_{\mu }=-\mathrm{i}\hbar \left( \,g_{\mu \,\nu }-\delta ^{2}{p_{\mu
}\,p_{\nu }}\right) \frac{\partial }{\partial p_{\nu }}+\mathrm{i}\hbar
\,\kappa \,p_{\mu }
\end{equation}

Now, in order that the position operators have physical sense, they should represent
physical observables, that is to say, they must be hermitian operators. It is not
difficult to verify that the operators $q_\mu$, are not hermitian with the
internal product usually employed in quantum mechanics:
\begin{equation}
\label{Eq19}
\left(\psi,\phi\right)=\int\,d\tau\,\psi^*\,\phi
\end{equation}
\noindent There are two possible alternatives that they are exactly
equivalent to each other:

\begin{enumerate}
\item Construct an internal product in which our position operators $q_{\mu } $
are hermitian with $\kappa$ arbitrary.

\item Choose the constant $\kappa $ in such a way that these  operators
are hermitian with the usual internal product.
\end{enumerate}

The second possibility is quite simple since it is enough to impose that
the operators ${q}_{\mu }$ are hermitian with the internal product ~(\ref{Eq19});
the result is:
\begin{equation}
\kappa =\frac{N+1}{2}\,\delta ^{2}
\end{equation}%
where $N=1,2,3,4$ is the dimension of the space time.

In this work, however, we will explore the first possibility. Consequently,
we needed to construct an internal product in which the operators $q_{\mu }$
are self adjoint. We postulate the general form:
\begin{equation}
\left( \psi ,\phi \right) =\int \,d\tau \,\frac{\psi ^{\ast }\,\phi }{%
\mathrm{W}\left( p\cdot p\right) }
\end{equation}%
where $d\tau =dp_{0}dp_{1}dp_{2}dp_{3}$, $\mathrm{W}\left( p\cdot p\right) $
is a weight function to be determined, and $p\cdot p=g_{\mu \,\nu }p_{\mu
}p_{\nu }$. Imposing condition of hermeticity of the operators $q_{\mu }$ with
this new internal product:
\begin{equation}
\label{Eq22}
\left( q_{\mu }\psi ,\phi \right) =\left( \psi ,q_{\mu }\phi \right)
\end{equation}%
and requiring which the functions $\psi $ and $\phi $ vanish suitably fast
at infinity, a partial integration shows that, to insure the fulfillment of
condition~(\ref{Eq22}), the weight function $\mathrm{W}$ satisfy the following differential equation:
\begin{equation}
\left( g_{\mu \,\nu }+\delta ^{2}p_{\mu }p_{\nu }\right) \frac{\partial
\mathrm{W}}{\partial \,p_{\nu }}+\left[ (N+1)\,\delta ^{2}-2\,\kappa \right]
\,p_{\mu }\,\mathrm{W}=0
\end{equation}%
where $N$ is the number of dimensions of the space time. The most general
solution for $\mathrm{W}$ is:
\begin{equation}
\label{Eq24}
\mathrm{W}=\left( 1-\delta ^{2}g_{\mu \,\nu }p_{\mu }p_{\nu }\right)
^{1-\beta }
\end{equation}%
where a multiplicative integration constant has been chosen equal to unity,
since any constant of this type may always be included in the normalization
of the wave functions $\psi $, $\phi $. The constant $\beta $ is given by:
\begin{equation}
\beta =\frac{\kappa }{\delta ^{2}}-\frac{1}{2}\,(N-1)
\end{equation}%
Finally, the sought-after internal product is:
\begin{equation}
\left( \psi ,\phi \right) =\int \,d\tau \,\frac{\psi ^{\ast }\,\phi }{\left(
1-\delta ^{2}\,g_{\mu \,\nu }p_{\mu }\,p_{\nu }\right) ^{1-\beta }}
\end{equation}%
from which we can define the probability amplitude:
\begin{equation}
{\Psi }(p)=\frac{\psi (p)}{\left( 1-\delta ^{2}g_{\mu \,\nu }p_{\mu
}\,p_{\nu }\right) ^{\frac{1-\beta }{2}}}
\end{equation}%
and the probability density:
\begin{equation}
\rho (p)=\Psi ^{\ast }(p)\Psi (p)=\frac{|\psi (p)|^{2}}{\left( 1-\delta
^{2}g_{\mu \,\nu }p_{\mu }\,p_{\nu }\right) ^{1-\beta }}
\end{equation}

\subsection{Algebraic properties of space time}

>From equation~(\ref{Eq18}) we have:
\begin{eqnarray}
\label{Eq29}
q_{k} &=&\mathrm{i}\hbar \left[ \frac{\partial }{\partial p_{k}}+\delta
^{2}p_{k}\left( p_{\nu }\frac{\partial }{\partial p_{\nu }}\right) \right] +%
\mathrm{i}\hbar \,\kappa \,p_{k}\quad k=1,2,3 \\
\label{Eq30}
q_{0} &=&-\mathrm{i}\hbar \left[ \frac{\partial }{\partial p_{0}}-\delta
^{2}p_{0}\left( p_{\nu }\frac{\partial }{\partial p_{\nu }}\right) \right] +%
\mathrm{i}\hbar \,\kappa \,p_{0}
\end{eqnarray}%
These operators look much like the operators introduced by Snyder in~\cite{HS2}, the
differences are in the additional terms $\kappa \,p_{k}$ and $\kappa \,p_{0}$
respectively, in addition, (\ref{Eq30}) has a global minus sign in the first term.

Now we define the operators:
\begin{equation}
L_j=\epsilon_{j\,k\,\ell}\,q_kp_\ell ,\quad M_k=q_kp_0-q_0p_k
\end{equation}
where $\epsilon_{j\, k\, \ell}$ is the usual Levi-Civita symbol.
Making some simple algebraic manipulations, we find that $L_j$ and $M_k$
have the same explicit expression as in the usual quantum mechanics:
\begin{equation}
L_j=-\mathrm{i}\hbar\,\epsilon_{j\,k\,\ell}\,p_k\,\frac{\partial}{\partial
p_\ell}\quad M_k=\mathrm{i}\hbar\,\left(p_0\,\frac{\partial}{\partial p_k}+
p_k\,\frac{\partial}{\partial p_0}\right)
\end{equation}
and they are the infinitesimal generators of the Lorentz group:
\begin{eqnarray}
\label{Eq33}
\left[L_j,L_k\right]&=&\mathrm{i}\hbar\epsilon_{j\,k\,\ell}\,L_\ell \\
\label{Eq34}
\left[M_j,M_k\right]&=&-\mathrm{i}\hbar\epsilon_{j\,k\,\ell}\,M_\ell \\
\label{Eq35}
\left[L_j,M_k\right]&=&\mathrm{i}\hbar\epsilon_{j\,k\,\ell}\,M_\ell
\end{eqnarray}
Evidently $L_j$ has the usual properties of the angular momentum in quantum
mechanics. In addition, other direct calculations allow us to show that:
\begin{eqnarray}
\left[q_j,q_k\right]&=&\mathrm{i}\hbar\,\delta^2\epsilon_{j\,k\,\ell}\,L_\ell
\\
\left[q_0,q_k\right]&=&\mathrm{i}\hbar\delta^2M_k \\
\left[q_j,L_k\right]&=&\mathrm{i}\hbar\,\epsilon_{j\,k\,\ell}\,q_\ell \\
\left[p_j,L_k\right]&=&\mathrm{i}\hbar\,\epsilon_{j\,k\,\ell}\,p_\ell
\end{eqnarray}
The position operators do not commute, their commutators are proportional to
the infinitesimal generators of the Lorentz group. Evidently:
\begin{eqnarray}
\Delta q_j\Delta q_k&\neq&\,0,\quad j,k=1,2,3 \\
\Delta q_0\Delta q_k&\neq&\,0,\quad k=1,2,3
\end{eqnarray}
That is to say, in this theory, it is not possible to measure two
coordinates simultaneously.

The algebra obtained when constructing the commutation relations~(\ref{Eq33}),
(\ref{Eq34}), and (\ref{Eq35}) is identical to the one of Snyder~\cite{HS1}. Consequently,
the proposed theory has the important property of being relativistically
invariant. This is not accidental, since the commutation relation~(\ref{Eq12}), from
which the theory is deduced, are evidently covariant. We will see below  that the
space time in this theory is discrete, however, we have already seen that Lorentz invariance
is included from the begining, it is a fundamental requirement here.

\subsection{Structure of the space time}

In order to find the nature of physical space, we must solve the eigenvalues equation:
\begin{equation}
\label{Eq42}
q_{\mu }\psi =\lambda \,\psi
\end{equation}%
with the operators $q_{\mu }$ given by ~(\ref{Eq18}). In one dimension the
equation ~(\ref{Eq42}), with $\kappa =0$, is:
\begin{equation}
\label{Eq43}
\mathrm{i}\hbar \left[ \frac{d}{dp}+\delta ^{2}p\left( p\frac{d}{dp}\right) %
\right] \psi (p)=\lambda \,\psi (p)
\end{equation}%
this equation can be solved immediately. Nevertheless, is interesting to
study it in an auxiliary space, to which we will call \textit{background
space}. In this space the commutation relationships are the usual in quantum
mechanics, that is to say:
\begin{eqnarray}
\left[ \hat{x}_{j},\hat{p}_{k}\right]  &=&\mathrm{i}\hbar \,\delta
_{j\,k},\qquad j,k=1,2,3 \\
\left[ {\hat{x}}_{0},{\hat{p}}_{0}\right]  &=&-\mathrm{i}\hbar
\end{eqnarray}%
We notice that the operator $q$ in (\ref{Eq43}) becomes the operator
$\hat{x}$ when $\delta \rightarrow 0$, hence the
name given to the space. Interpreting equation ~(\ref{Eq42}) like a
quantum equation in the momentum representation:
\begin{equation}
\left[ \hat{x}+\delta ^{2}\hat{p}\left( \hat{p}\hat{x}\right) \right] \psi
(p)=\lambda \,\psi (p)
\end{equation}%
we can write, in \textit{background representation }:
\begin{equation}
\label{Eq47}
\left[ x-\hbar ^{2}\delta ^{2}\frac{d}{dx}\left( \frac{d}{dx}\,x\right) %
\right] {\tilde{\psi}}(x)=\lambda \,{\tilde{\psi}}(x)
\end{equation}%
where ${\tilde{\psi}}(x)$ is the Fourier transform of ${\psi }(p)$. This
shows that we can solve the eigenvalue problem of the position operator, either
in the momentum representation, or in the coordinate representation provided by
\textit{the background space }. This this last one is obtained by the following transformations:
\begin{equation}
\label{Eq48}
p_{k}\!\rightarrow \!-\mathrm{i}\hbar \frac{\partial }{\partial x_{k}},\
\frac{\partial }{\partial p_{k}}\!\rightarrow \!-\frac{\mathrm{i}}{\hbar }%
x_{k},\ p_{0}\!\rightarrow \!\mathrm{i}\hbar \frac{\partial }{\partial x_{0}}%
,\ \frac{\partial }{\partial p_{0}}\!\rightarrow \!\frac{\mathrm{i}}{\hbar }%
x_{0},\ \psi \!\rightarrow \!\tilde{\psi}
\end{equation}%
As an example, writing:
\begin{equation}
\eta =\frac{2\,x}{\hbar \,\delta },\quad \tilde{\psi}=A\,e^{{-\frac{\eta }{2}%
}}\,\phi (\eta )
\end{equation}%
equation ~(\ref{Eq47})  is reduced to:
\begin{equation}
\eta \,\frac{d^{2}\phi }{d\,\eta ^{2}}+(1+1-\eta )\,\frac{d\phi }{d\eta }%
+\left( \frac{\lambda }{2\,\hbar \,\delta }-1\right) \,\phi =0
\end{equation}%
the solution, therefore, are the usual associate Laguerre polynomials:
\begin{equation}
\phi (\eta )=L_{n-1}^{1}(\eta ),\quad n\geq 1
\end{equation}%
with eigenvalues:
\begin{equation}
\lambda _{n}=2\,n\,\hbar \,\delta
\end{equation}%
This result, is the same as the one obtained, by a different procedure, in
reference~\cite{Saavedra3}. Using the rules given in equation~(\ref{Eq48}), it may be
easily verified that the eigenvalues $\lambda _{n}$ of the position
operator $q$ do not depend on the choice of the constant $\kappa $.

When the previous procedure is applied to the operators $q_{k}$ and $q_{0}$
gives by the equations ~(\ref{Eq29}) and ~(\ref{Eq30}) we obtain the following set of
operators in the background space:
\begin{eqnarray}
\label{Eq53}
\zeta _{k} &=&x_{k}+\hbar ^{2}\delta ^{2}\frac{\partial }{\partial x_{k}}
\left( x_{\nu }\,\frac{\partial }{\partial x_{\nu }}\right) +\hbar
^{2}\left( \kappa -N\,\delta ^{2}\right) \,\frac{\partial }{\partial x_{k}}
\\
\label{Eq54}
\zeta _{0} &=&x_{0}+\hbar ^{2}\delta ^{2}\frac{\partial }{\partial x_{0}}%
\left( x_{\nu }\,\frac{\partial }{\partial x_{\nu }}\right) +\hbar
^{2}\left( \kappa -N\,\delta ^{2}\right) \,\frac{\partial }{\partial x_{0}}
\end{eqnarray}%
which have properties similar to those of the operators $q_{k}$ and
$q_{0}$, and satisfy an identical algebra. This is correct, of
course, since both operators sets are related by a unitary
transformation.

Using the fact that $\kappa $ is an arbitrary parameter we can establish relationships
with another result existing in the literature. In fact, setting $\kappa =N\,\delta ^{2}$,
the operators $\zeta _{k},\zeta _{0}$ given in~(\ref{Eq53}) and (\ref{Eq54}) are reduced to
the set of operators introduced by Hellund and Tanaka~\cite{Hellund} to describe a
quantized space time. These authors assume that time is a continuous variable and
demonstrate that the operators $\zeta _{k}$ have discrete spectrum.

In our formalism it is possible to demonstrate explicitly that the operator $q_{0}$
has continuous spectrum, that is to say, in this theory, the time is a
continuous variable. The eigenvalue equation for the operator $q_{0}$
given in ~(\ref{Eq30}) is:
\begin{equation}
q_{0}\psi =\lambda \psi
\end{equation}%
and its solution is:
\begin{equation}
\label{Eq56}
\psi =A\,\frac{\exp \left[ \frac{\mathrm{i}\lambda }{\hbar \delta }\,\tanh
^{-1}\left( \delta \,p_{0}\right) \right] }{\left( 1-\delta ^{2}\,g_{\mu
\,\nu }\,p_{\mu }\,p_{\nu }\right) ^{\frac{\kappa }{2\,\delta ^{2}}}}
\end{equation}%
where $A$ it is a normalization constant. Integrating completely the spatial
part we obtain:
\begin{equation}
|\psi |^{2}=\frac{4\,\pi }{3\,\delta ^{3}}\,|A|^{2}\,\int \frac{d\,p_{0}}{1-\delta ^{2}\,p_{0}^{2}}
\end{equation}%
Evidently, the integrand is singular **in one and minus one**. Now, to insure
that the theory has physical meaning, $\ell $ must be the smaller length that it
may appears in the problem. In particular, $\ell $ must be smaller than the
Compton wavelength, $\lambda _{c}$, of the particle of mass $m$, that is
to say, $\ell /\lambda _{c}<1$, and therefore $-1\leq \delta \,p_{0}\leq 1$.
In addition, with an appropriate choice of the constant $A$ it is possible
to prove directly that:

\begin{equation}
\left( \psi _{1},\psi _{2}\right) =\delta \left( \lambda _{1}-\lambda
_{2}\right)
\end{equation}%
where $\delta \left( \lambda _{1}-\lambda _{2}\right) $ is the Dirac delta
function and $\psi $ this given by $(56)$. The spectrum of $q_{0}$ is
continuous.

The generalization of the commutation relation proposed in equation ~(\ref{Eq12})
leads to a relativistic model where the space is discrete, but time is
continuous, allowing in this way the validity of the Heisenberg equation of
motion, as we assumed from the beginning.

\section{Energy spectra}

We have seen already that the commutation relation ~(\ref{Eq12}) determines the
position operators in the form: $q_{\mu }=q_{\mu }\left( p,\frac{\partial }
{\partial p}\right) $. The position operators of this theory depend only on
the momentum. In principle, this allows us to solve the eigenvalue problem
for the Hamiltonian of any quantum system, in the momentum
representation. In this section we shall calculate the energy spectrum and
the eigenfunctions of the one dimensional harmonic oscillator and the
three-dimensional isotropic harmonic oscillator. That is to say, we will
solve in both cases the eigenvalues equation:
\begin{equation}
\label{Eq59}
H\,\Psi =E\,\Psi
\end{equation}%
introducing in $H=H(q,p)$ the position operators ~(\ref{Eq29}), which in the one
dimensional case become:
\begin{equation}
\label{Eq60}
q=\mathrm{i}\hbar \left( 1+\delta ^{2}\,p^{2}\right) \frac{\partial }{%
\partial \,p}+\mathrm{i}\hbar \,\kappa \,p
\end{equation}%
In the three-dimensional case we will use:
\begin{equation}
\label{Eq61}
q_{k}=\mathrm{i}\hbar \left[ \frac{\partial }{\partial p_{k}}+\delta
^{2}p_{k}\left( p_{\mathrm{j}}\frac{\partial }{\partial p_{\mathrm{j}}}%
\right) \right] +\mathrm{i}\hbar \,\kappa \,p_{k}\quad \mathrm{j},k=1,2,3
\end{equation}%
We remark that the examples that we present here are clearly non
relativistic. However, they are valid examples because in the non
relativistic case our generalization becomes rotationally invariant.

\subsection{The Harmonic Oscillator in one Dimension}

The Hamiltonian of the system is:
\begin{equation}
H=\frac{p^{2}}{2\,m}+\frac{1}{2}\,m\,\omega ^{2}\,q^{2}
\end{equation}%
Replacing $q$ by ~(\ref{Eq60}) and rearranging, the equation of eigenvalues ~(\ref{Eq59})
becomes:
\begin{eqnarray}
&-&\frac{1}{2}m\hbar ^{2}\omega ^{2}\left( 1+\delta ^{2}p^{2}\right) ^{2}%
\frac{d^{2}\Psi }{dp^{2}}-m\hbar ^{2}\omega ^{2}\left( \kappa +\delta
^{2}\right) \left( 1+\delta ^{2}p^{2}\right) \frac{d\Psi }{dp}  \nonumber \\
&-&\left[ E-\frac{p^{2}}{2\,m}+\frac{1}{2}\,m\,\omega ^{2}\left( 1+(\kappa
+\delta ^{2}\,)\,p^{2}\right) \right] \Psi =0
\end{eqnarray}%
Introducing the changes:
\begin{equation}
p=\sqrt{m\,\hbar \,\omega }\,P,\qquad \Psi (p)\rightarrow \Psi (P)
\end{equation}%
and using the given definition of $\delta $ ~(\ref{Eq13}) explicitly we obtain:
\begin{eqnarray}
&{}&\left( 1+\frac{\omega \,\ell }{c}\,P^{2}\right) ^{2}\frac{d^{2}\Psi }{%
dP^{2}}+2\,m\,\hbar \omega \left( \kappa +\frac{\ell }{m\,\hbar \,c}\right)
\left( 1+\frac{\omega \,\ell }{c}\,P^{2}\right) P\frac{d\Psi }{dP}  \nonumber
\\
&+&\left[ \frac{2E}{\hbar \omega }+\kappa m\hbar \omega +\left( \kappa
m\hbar \omega \frac{\omega \ell }{c}+\kappa ^{2}m^{2}\hbar ^{2}\omega
^{2}-1\right) P^{2}\right] \Psi =0
\end{eqnarray}%
Now we will assume that the solutions of this differential equation have
the form:
\begin{equation}
\Psi =\left( 1+\xi ^{2}\right) ^{\sigma }\,\Phi (\xi )
\end{equation}%
where $\sigma $ is a unknown parameter that we will fix by imposing that the
resulting differential equation for $\Phi (\xi )$ reduces to the Hermite
equation when we take the limit $\ell \rightarrow 0$. The detailed calculations
are actually quite simple, and two possible values for $\sigma $ are obtained:
\begin{eqnarray}
\sigma _{1} &=&-\frac{1}{4}-\kappa \frac{m\hbar c}{2\ell }+\frac{1}{4}\,%
\sqrt{1+\left( \frac{2c}{\omega \ell }\right) ^{2}} \\
\sigma _{2} &=&-\frac{1}{4}-\kappa \frac{m\hbar c}{2\ell }-\frac{1}{4}\,%
\sqrt{1+\left( \frac{2c}{\omega \ell }\right) ^{2}}
\end{eqnarray}%
For reasons of consistency with the known results of the usual quantum
mechanics, we will choose $\sigma =\sigma _{2}$. The resulting equation for
$\Phi $ is:
\begin{eqnarray}
&{}&\left( 1\!+\!\frac{\omega \,\ell }{c}\,P^{2}\right) ^{2}\!\frac{%
d^{2}\Phi }{dP^{2}}\!+\!\frac{\omega \ell }{c}\left( 1\!-\!\sqrt{%
1\!+\!\left( \frac{2\,c}{\omega \ell }\right) ^{2}}\,\right) \left( 1\!+\!%
\frac{\omega \,\ell }{c}\,P^{2}\right) \!P\frac{d\Phi }{dP}  \nonumber \\
&+&\left[ \frac{2E}{\hbar \omega }-\frac{\omega \ell }{2\,c}\left( 1\!+\!%
\sqrt{1\!+\!\left( \frac{2\,c}{\omega \ell }\right) ^{2}}\,\right) \right]
\Phi =0
\end{eqnarray}%
Finally, making the changes:
\begin{equation}
P=\sqrt{\frac{c}{\omega \ell }}\,\frac{\eta }{\sqrt{1-\eta ^{2}}}\,,\quad
-1\leq \eta \leq 1\,\quad \Phi (P)\rightarrow \Phi (\eta )
\end{equation}%
the following differential equation is obtained:
\begin{eqnarray}
&{}&\left( 1-\eta ^{2}\right) \frac{d^{2}\Phi }{d\eta ^{2}}-\left( 2\!+\!%
\sqrt{1\!+\!\left( \frac{2\,c}{\omega \ell }\right) ^{2}}\,\right) \eta
\frac{d\Phi }{d\eta }  \nonumber \\
\label{Eq71}
&+&\left[ \frac{2\,c}{\omega \ell }\frac{E}{\hbar \omega }-\frac{1}{2}\left(
1\!+\!\sqrt{1\!+\!\left( \frac{2\,c}{\omega \ell }\right) ^{2}}\,\right) %
\right] \Phi =0
\end{eqnarray}%
Requiring that:
\begin{eqnarray}
\label{Eq72}
\frac{1}{2}+\frac{1}{2}\,\sqrt{1+\left( \frac{2\,c}{\omega \ell }\right) ^{2}%
} &=&a \\
\label{Eq73}
\frac{2\,c}{\omega \ell }\frac{E}{\hbar \omega }-\frac{1}{2}-\frac{1}{2}\,%
\sqrt{1+\left( \frac{2\,c}{\omega \ell }\right) ^{2}} &=&n(n+2\,a)
\end{eqnarray}%
where $n$ is an integer, the equation ~(\ref{Eq71}) can be made to agree with the
differential equation for the Gegenbauer polynomials~\cite{Murphy}:
\begin{equation}
\left( 1-\eta ^{2}\right) \frac{d^{2}\Phi }{d\eta ^{2}}-(1+2\,a)\eta \frac{%
d\Phi }{d\eta }+n(n+2\,a)\Phi =0
\end{equation}%
Using (\ref{Eq72}) and (\ref{Eq73}), the eigenvalues for the energy are obtained:
\begin{equation}
\label{Eq75}
E_{n}=\hbar \omega \left[ \left( n+\frac{1}{2}\right) \,\sqrt{1+\left( \frac{%
\omega \ell }{2\,c}\right) ^{2}}+\left( n^{2}+n+\frac{1}{2}\right) \frac{%
\omega \ell }{2\,c}\right]
\end{equation}%
Making the inverse change of variables, it is possible to demonstrate very
easily that the normalized eigenfunctions, in terms of the original
variables, are given by~\cite{Abramowitz}:
\begin{equation}
\Psi _{n}(p)=2^{a}\Gamma (a)\sqrt{\frac{n!(n+a)\delta }{2\pi \Gamma (n+2a)}}%
\left( 1+\delta ^{2}p^{2}\right) ^{\sigma }C_{n}^{a}\left( \frac{\delta p}{%
\sqrt{1+\delta ^{2}p^{2}}}\right)
\end{equation}%
where $C_{n}^{a}$ is the standard notation for the Gegenbauer polynomials.

\subsection{The three-dimensional isotropic harmonic oscillator}

The Hamiltonian of this system:
\begin{equation}
\label{Eq77}
H=\frac{1}{2\,m}\,p_{k}\,p_{k}+\frac{1}{2}\,m\,\omega
^{2}\,q_{k}\,q_{k}\,,\qquad k=1,2,1
\end{equation}%
is rotationally invariant. Then, using spherical coordinates $(p,\vartheta
,\varphi )$, in the momentum space and replacing $q_{k}$ by ~(\ref{Eq61}) we find
that the eigenvalue equation ~(\ref{Eq59}, in this case, takes the form:
\begin{eqnarray}
&{}&\frac{p^{2}}{2\,m}\,\Psi -\frac{1}{2}\,m\,\hbar ^{2}\omega ^{2}\,\left(
1+\delta ^{2}p^{2}\right) ^{2}\frac{\partial ^{2}\Psi }{\partial p^{2}}
\nonumber \\
&-&\frac{1}{2}\,m\,\hbar ^{2}\,\omega ^{2}\,\left( 1+\delta ^{2}p^{3}\right) %
\left[ 1+\left( \kappa +\delta ^{0}\right) p^{2}\right] \frac{2}{p}\frac{%
\partial \Psi }{\partial p}  \nonumber \\
&+&\frac{1}{2}\,m\,\omega ^{2}\frac{L^{2}}{p^{2}}\,\Psi -\frac{1}{2}\,\kappa
\,m\,\hbar ^{2}\,\omega ^{2}\,\left[ 3+\left( \kappa +\delta ^{2}\right)
p^{2}\right] \Psi =E\,\Psi
\end{eqnarray}%
where $L$ is the usual angular momentum operator in quantum mechanics. Then,
writing the eigenfunctions of the Hamiltonian $H$ like a product of the
spherical harmonic $\mathrm{Y}_{s\,m}(\vartheta ,\varphi )$ and one radial
function $\Pi (p)$~\cite{Sommerfeld}, we obtain:\footnote{$L^{2}\mathrm{Y}
_{s\,m}(\vartheta ,\varphi )=\hbar ^{2}\,s(s+1)\,\mathrm{Y}_{s\,m}(\vartheta
,\varphi ),\quad s=|m|$}
\begin{eqnarray}
&{}&\left( 1+\delta ^{2}p^{2}\right) ^{2}\frac{d^{2}\Pi }{dp^{2}}+\left(
1+\delta ^{2}p^{2}\right) \left[ 1+\left( \kappa +\delta ^{2}\right) p^{2}
\right] \frac{2}{p}\frac{d\Pi }{dp}  \nonumber \\
&+&\left[ \!\frac{2E}{m\hbar ^{2}\omega ^{2}}\!-\!\frac{p^{2}}{m^{2}\hbar
^{2}\omega ^{2}}\!-\!\frac{s(s\!+\!1)}{p^{2}}\!+\!\kappa \left[ 3\!+\!\left(
\kappa \!+\!\delta ^{2}\right) p^{2}\!\right] \!\right] \!\Pi \!=\!0
\end{eqnarray}%
As it is usual, to remove the singularity when $p\rightarrow 0$ we do:
\begin{equation}
\Pi (p)=p^{s}\,\Phi (p)
\end{equation}%
In the resulting equation for $\Phi (p)$ we make a new change:
\begin{equation}
\Phi (p)=\left( 1+\delta ^{2}p^{2}\right) ^{\gamma }\Xi (p)
\end{equation}%
and we pick $\gamma $ so that the singularity when $p\rightarrow \infty $
disappears. Like in the one dimensional problem this provides two possible
choices for $\gamma $. We chose the one that gives us the correct
limits fot the spectrum and the eigenfunctions when we let $\ell
\rightarrow 0$:
\begin{equation}
\gamma =-\frac{1}{4}-\frac{s}{2}-\kappa \,\frac{m\,\hbar \,c}{2\,\ell }-%
\frac{1}{4}\,\sqrt{1+\left( \frac{2\,c}{\omega \ell }\right) ^{2}}
\end{equation}%
Finally, in the resulting equation for $\Xi (p)$, we make following
changes:
\begin{equation}
\label{Eq83}
p=\frac{1}{\delta }\,\sqrt{\frac{1+\eta }{1-\eta }}\,,\quad -1\leq \eta \leq
1\,,\quad \Xi (p)\rightarrow \Xi (\eta )
\end{equation}%
The result of all these operations is the following differential equation
for $\Xi (\eta )$:
\begin{eqnarray}
&{}&\left( 1-\eta ^{2}\right) \frac{d^{2}\Xi }{d\eta ^{2}}  \nonumber \\
&+&\left[ s+\frac{1}{2}-\frac{1}{2}\,\sqrt{1+\left( \frac{2\,c}{\omega \ell }%
\right) ^{2}}-\left( \frac{5}{2}+s+\frac{1}{2}\,\sqrt{1+\left( \frac{2\,c}{%
\omega \ell }\right) ^{2}}\right) \eta \right] \frac{d\Xi }{d\eta }
\nonumber \\
&+&\left[ \frac{c}{2\omega \ell }\frac{E}{\hbar \omega }-\frac{3}{8}-\frac{s%
}{2}-\left( \frac{3}{8}+\frac{s}{4}\right) \sqrt{1+\left( \frac{2\,c}{\omega
\ell }\right) ^{2}}\right] \Xi =0
\label{Eq84}
\end{eqnarray}%
where we have used explicitly the given definition of $\delta $ (\ref{Eq13}). Let

\begin{eqnarray}
\label{Eq85}
\frac{1}{2}\,\sqrt{1+\left( \frac{2\,c}{\omega \ell }\right) ^{2}} &=&a \\
\label{Eq86}
s+\frac{1}{2} &=&b \\
\label{Eq87}
\frac{c}{2\omega \ell }\frac{E}{\hbar \omega }\!-\!\frac{3}{8}\!-\!\frac{s}{2%
}\!-\!\left( \frac{3}{8}\!+\!\frac{s}{4}\right) \sqrt{1\!+\!\left( \frac{2\,c%
}{\omega \ell }\right) ^{2}} &=&m^{\prime }(m^{\prime }\!+\!a\!+\!b\!+\!1)
\end{eqnarray}%
The equation (\ref{Eq84}) can be made to agree with the differential equation for the
Jacobi polynomials~\cite{Abramowitz,Gradshteyn}:
\begin{equation}
\label{Eq88}
\left( 1\!-\!\eta ^{2}\right) \frac{d^{2}\Xi }{d\eta ^{2}}+\left[
b\!-\!a\!+\!(a\!+\!b\!+\!2)\eta \right] \frac{d\Xi }{d\eta }+m^{\prime
}(m^{\prime }\!+\!a\!+\!b\!+\!1)\Xi \!=\!0
\end{equation}%
Using ~(\ref{Eq85}-\ref{Eq87}) we can obtain the eigen-energies of the three-dimensional
isotropic harmonic oscillator
\begin{equation}
\label{Eq89}
E_{n}\!=\!\hbar \omega \left( n\!+\!\frac{3}{2}\right) \sqrt{1\!+\!\left(
\frac{\omega \ell }{2\,c}\right) ^{2}}\!+\!\hbar \omega \left(
n^{2}\!+\!3\,n-s(s+1)\!+\!\frac{3}{2}\right) \frac{\omega \ell }{2\,c}
\end{equation}%
where $n=s+2\,m^{\prime }$. The solution of ~(\ref{Eq88}) are the so called Jacobi
polynomials, $\Xi (\eta )=P_{n}^{(a,b)}(\eta )$.

The eigenfunctions of Hamiltonian ~(\ref{Eq77}) are:
\begin{equation}
\Psi _{smn}(p,\vartheta ,\varphi )=\mathrm{Y}_{s\,m}(\vartheta ,\varphi
)\,A\,\Pi _{n}(p)
\end{equation}%
where $A$ is a normalization constant. Choosing the spherical harmonical
properly normalized:
\begin{equation}
\mathrm{Y}_{s\,m}(\vartheta ,\varphi )=\sqrt{\frac{2\,s+1}{4\,\pi }\frac{%
(s-m)!}{(s+m)!}}\,P_{s}^{m}(\cos \theta )e^{\mathrm{i}\varphi }
\end{equation}%
where $P_{s}^{m}(\cos \theta )$ are the associated Legendre polynomials, the
constant $A$ is completely determined by the normalization condition:
\begin{equation}
\left\vert A\right\vert ^{2}\int_{0}^{\infty }\Pi _{n}(p)\Pi _{n}^{\ast }(p)%
\frac{p^{2}\,dp}{W}=1
\end{equation}%
where $W$ is the weight function ~(\ref{Eq24}). Replacing $\Pi _{n}(p)$ and using
the change of variables ~(\ref{Eq83}) we obtain exactly the normalization integral
for the Jacobi polynomials:
\begin{equation}
\left\vert A\right\vert ^{2}\,\frac{2^{2\,\gamma +\beta -1}}{\delta ^{2\,s+3}%
}\,\int_{-1}^{1}(1-\eta )^{a}(1+\eta )^{b}\left\vert P_{n}^{(a,b)}(\eta
)\right\vert ^{2}\,d\eta =1
\end{equation}%
where all the symbols used have been previously defined. The value of the
integral is very well known~\cite{Gradshteyn}, and we obtain $A$. Returning to the
original variables, the normalized radial function is:
\begin{eqnarray}
\Pi _{n}(p)={} &&\sqrt{\frac{\,2\,(2n+a+b+1)\,n!\,\Gamma (n+a+b+1)}{\Gamma
(n+a+1)\Gamma (n+b+1)}}\,\delta ^{3/2}  \nonumber \\
&{}&\times \left( \delta \,p\right) ^{s}\,\left( 1+\delta ^{2}\,p^{2}\right)
^{\gamma }P_{n}^{(a,b)}\left( \frac{\delta ^{2}p^{2}-1%
\ddot{} }{\delta ^{2}p^{2}+1}\right)
\end{eqnarray}%
where $P_{n}^{(a,b)}$ is the standard notation for the Jacobi polynomials.

\section{Summary and Conclusions}

In this work we have presented a relativistic generalization of the
canonical commutation relation ~(\ref{Eq01}).

The main consequence is that, within this theory, the physical space becomes a
discrete set of points, whereas the time variable is continuous, as in the usual theory.
This result is in agreement with the initial hypothesis, that within this
generalization, the Heisenberg equation of motion is valid.

The position operators $q_\mu$ satisfy an algebra that is formally identical
to algebra of Snyder~\cite{HS1}. Their commutators are proportional to the
infinitesimal generators of the Lorentz group: $\Delta q_j\Delta q_k\neq\,
0, \ \Delta q_0\Delta q_k\neq\, 0, \ j, k=1, 2, 3$. Due to this, it is not
possible to measure simultaneously two coordinates within this theory.

In addition, within this formalism, the product of the incertitudes of the
position and momentum operators $\Delta q_{\mu }\Delta p_{\nu }\neq \,
0, \forall \ \mu, \ \nu $. In particular, for the one dimensional case,
$ \Delta q\, \Delta p\propto p^{2}$. In the low energy regime, when the
mass is great compared with the momentum, this result reduces to $\Delta
q\, \Delta p\propto E$.

We also showed that an intermediate representation of the position operators
exists, in the sense that it is not the representation of momentum nor the
representation of coordinates; we named it \textit{background space} instead. In
this representation, the eigenvalue problem for the position operators
may be solved in terms of differential equations very well known in
quantum mechanics.

Finally, we have calculated the energy spectrum of two simple
quantum systems: \textit{the one dimensional harmonic oscillator}
and \textit{the three-dimensional isotropic harmonic oscillator}. In
both systems the energy levels depend on $n^{2}$. This fact without
a doubt is a reflection of our modified commutation relation. The
results ~(\ref{Eq75}) and ~(\ref{Eq89}) show that although the non-dimensional
parameter: $\omega \,\ell /2\,c$ may be small, the deviation of the
usual dependency in $n$ will be pronounced for sufficiently great
values of $n$.

\ack {It is a great pleasure to thank Professor Dr. Igor Saavedra,
who suggested this problem a long time ago. H. Calisto acknoledges finantial
support from Universidad de Tarapac\'a ( UTA Grant \# 4723-05). C.A. Utreras-D\'{i}az 
acknowledges support from Universidad Austral de Chile (DID Grant \#
S-2004-43), and FONDECYT (Grant \#1040311).

\end{document}